\documentclass[twocolumn,pra,showpacs,superscriptaddress,amssymb,amsmath,amsmath]{revtex4-1}
\usepackage{graphicx}
\usepackage{epstopdf}
\usepackage{bm}
\usepackage{hyperref}
\usepackage{comment}
\usepackage{color}
\usepackage{physics}
\usepackage{amsmath}
\usepackage{tikz}
\usepackage{enumitem}

\hypersetup{%
   pdfpagemode=None, %FullScreen,
   pdfstartpage=1,
   pdfmenubar=true,
   pdftoolbar=true,
   colorlinks = true,
   linkcolor=blue,
   citecolor=blue,
   urlcolor=blue,
   bookmarksopen=false
 }

\newcommand{\be}{\begin{equation}}
\newcommand{\ee}{\end{equation}}

\newcommand*\circled[1]{\tikz[baseline=(char.base)]{
            \node[shape=circle,draw,inner sep=2pt] (char) {#1};}}

\begin{document}

\title{Magnetic properties and quench dynamics of two interacting ultracold molecules \\in a trap}

\author{Anna Dawid}
\email{anna.dawid@fuw.edu.pl}
\affiliation{Faculty of Physics,  University of Warsaw, Pasteura 5, 02-093 Warsaw, Poland}
\affiliation{ICFO - Institut de Ci\`encies Fot\`oniques, The Barcelona Institute of Science and Technology, Av. Carl Friedrich Gauss 3, 08860 Castelldefels (Barcelona), Spain}
\author{Micha\l~Tomza}
\email{michal.tomza@fuw.edu.pl}
\affiliation{Faculty of Physics,  University of Warsaw, Pasteura 5, 02-093 Warsaw, Poland}

\date{\today}

\begin{abstract}

We theoretically investigate the magnetic properties and nonequilibrium dynamics of two interacting ultracold polar and paramagnetic molecules in a one-dimensional harmonic trap in external electric and magnetic fields. The molecules interact via a multichannel two-body contact potential, incorporating the short-range anisotropy of intermolecular interactions. We show that various magnetization states arise from the interplay of the molecular interactions, electronic spins, dipole moments, rotational structures, external fields, and spin-rotation coupling. The rich magnetization diagrams depend primarily on the anisotropy of the intermolecular interaction and the spin-rotation coupling. These specific molecular properties are challenging to calculate or measure. Therefore, we propose the quench dynamics experiments for extracting them from observing the time evolution of the analyzed system. Our results indicate the possibility of controlling the molecular few-body magnetization with the external electric field and pave the way towards studying the magnetization of ultracold molecules trapped in optical tweezers or optical lattices and their application in quantum simulation of molecular multichannel many-body Hamiltonians and quantum information storing.

\end{abstract}


\maketitle

%------------------------------------------------------------------------------
\section{Introduction}
%------------------------------------------------------------------------------
In the last two decades, experiments with ultracold atoms in optical lattices have provided tools allowing for unprecedented control and detection of ultracold quantum many-body systems~\cite{Bakr09Nature, Sherson10Nature, Serwane11, Boll16Science, EndresScience16, BarredoScience16} and resulted in successful quantum simulations of quantum many-body Hamiltonians of increasing complexity~\cite{Greiner02a, Greiner02b, Fukuhara13a, Fukuhara13b, Choi16, Mazurenko17, Bernien17, Chiu19}. These successes have been achieved even though atomic gases are typically governed by relatively simple isotropic and short-range interatomic interactions that are well described by the contact interaction~\cite{BlochRMP08}. Replacing atoms with molecules opens new possibilities resulting from the molecular rich internal structure, complex short-range interactions, and stronger long-range and anisotropic dipolar interactions~\cite{CarrNJP09,LemeshkoMP13,BohnScience17}.

A molecules' rich internal structure has earned them a prominent role in the precision measurements of fundamental constants~\cite{Kozlov02, Hudson02, Veldhoven04, Zelevinsky08, Bethlem09, Hudson11, Hutzler12, Baron14, Lim18}, while intermolecular dipolar interactions promise exciting applications in quantum information processing~\cite{Soderberg09, Zhu13, Park17, Ni18, Hughes20}. Ultracold molecules have also been employed in the ground-breaking experiments on quantum-controlled chemistry~\cite{Ospelkaus10, Ni10, Miranda11, TomzaPRL15, Klein17, Puri17, Jongh20} enabled by the extensive control of internal states and relative motion of molecules with external electromagnetic fields \cite{Friedrich00PCCP, Krems2005,Tscherbul06PRL, Tscherbul06JCP, Abrahamsson2007,Krems2008,Tscherbul09NJP,Alyabyshev09PRA,Lemeshko2013,Janssen13PRL}. Numerous applications of ultracold molecules in quantum simulations have been introduced, with a particular interest in studying quantum magnetism~\cite{Wall2015}. Molecular rotational states (in which pseudo-spins can be encoded with microwave-field dressing) combined with the dipolar interaction have allowed for several proposals to realize various models of quantum magnetism~\cite{MicheliNatPhys06, BarnettPRL06, GorshkovPRL11, Gorshkov11, HazzardPRL13, Manmana13}. Other theoretical works have concerned molecular quantum simulations of polarons~\cite{Ortner09, Herrera13}, rotating polarons~\cite{Schmidt15, Schmidt16}, magnetic Frenkel exciton \cite{Perez-Rios10NJP}, topologically nontrivial states~\cite{MicheliNatPhys06}, and exotic phases such as supersolid~\cite{Goral02, Wallis09PRL}.

A more complex internal structure of molecules, as compared to atoms, is responsible for greater experimental difficulties in molecular formation, cooling, and trapping. Despite these challenges, several species of ultracold molecules in their ground states have been produced via association of ultracold atoms~\cite{Ni08, Molony14, TakekoshiPRL14, Park15, He20Science} or recently using direct laser cooling from higher temperatures~\cite{CollopyPRL18,AndereggNP18}. Ultracold ground-state molecules have also been loaded into optical lattices~\cite{ChotiaPRL12}, and dipolar spin-exchange interactions between lattice-confined polar molecules have been observed~\cite{BoNature13}, opening the way towards quantum simulations with molecules. On the other hand, the methods of full quantum control, deterministic preparation, and detection at the single-particle level, developed for ultracold atoms in optical tweezers~\cite{EndresScience16,BarredoScience16}, can readily be employed to molecules~\cite{Liu18Science, Anderegg19Science}, further extending the range of applications of ultracold molecules~\cite{Cheuk20PRL}.

Precise spectroscopic characterization of molecular few-body systems of increasing complexity and reconstruction of underlying Hamiltonians may be both experimentally and theoretically challenging. One of the possible solutions may be quench dynamics experiments~\cite{GarciaMarch16, Usui18, Li20}. In such a scenario, the system parameters, such as the interparticle interaction strength, trapping potential or external fields, are suddenly changed~\cite{Mitra18}. This excites the system from the ground state and induces its time evolution, whose observation may reveal the system's intrinsic properties. The quench dynamics has been thoroughly studied for two \cite{Kehrberger18PRA, Budewig19MolPhys, Bougas19PRA, Guan19PRL} and three \cite{GarciaMarch16} ultracold atoms in a trap. Quantum quenches also allow one to study nonequilibrium dynamics~\cite{PolkovnikovRMP11,Gring12,ErneNat18}.

\begin{figure}[t]
\begin{center}
\includegraphics[width=\columnwidth]{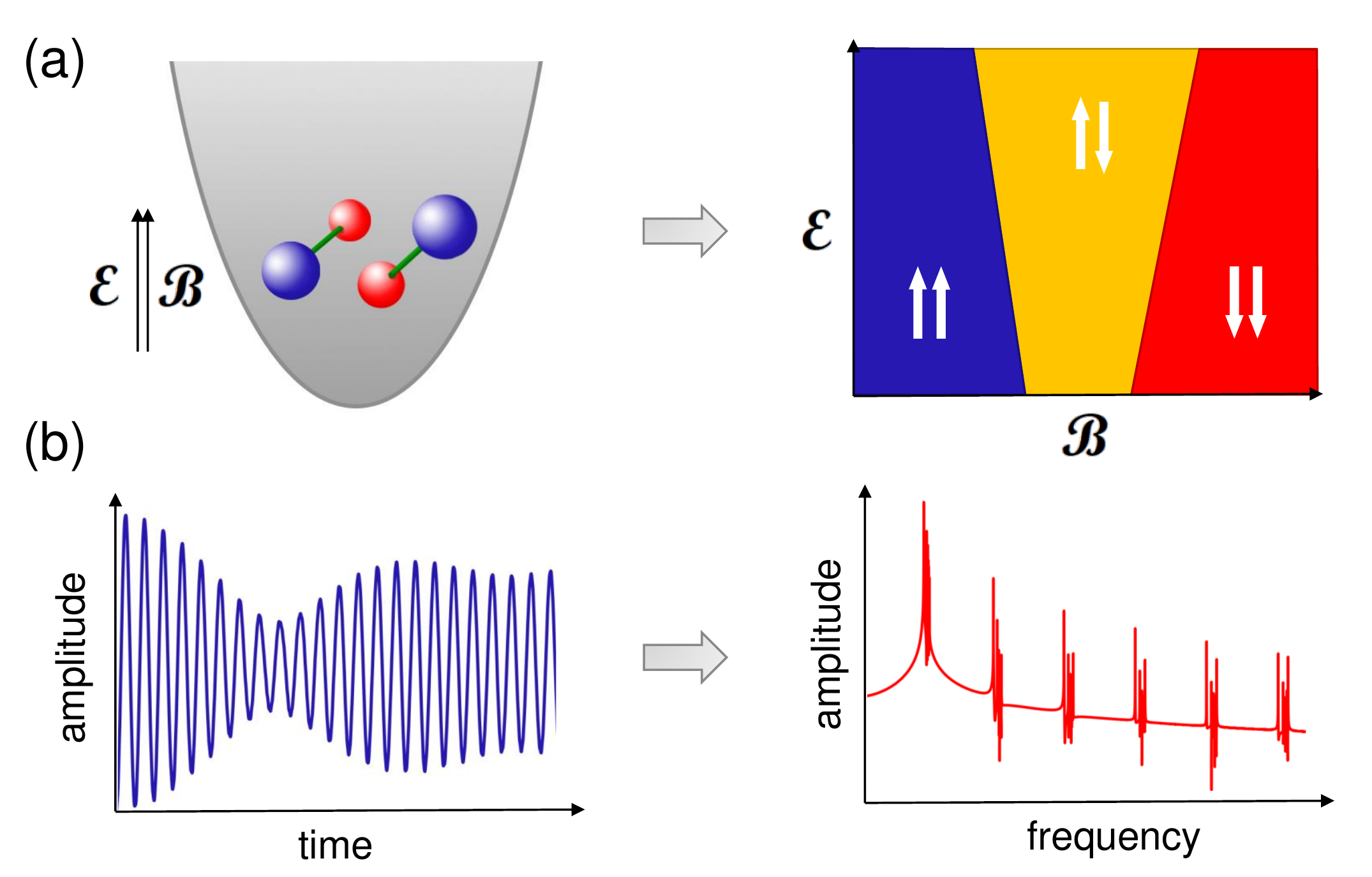}
\end{center}
\caption{Schematic representation of the investigated system and its features. (a) Two interacting molecules in a one-dimensional harmonic trap under the influence of the electric and magnetic fields can be described by a magnetization diagram depending on the field strengths. The fields are parallel to each other and to the direction of molecular motion. (b) Time evolution of the system's observable after the quench can reveal information on the underlying molecular characteristics by using the Fourier transform.}
\label{fig:intro}
\end{figure}

In this work, we study the magnetic properties and nonequilibrium dynamics of two interacting ultracold polar and paramagnetic molecules in a one-dimensional harmonic trap (see Fig.~\ref{fig:intro}). Intriguing features of the system arise from the interplay of the molecular electronic spins, dipole moments, rotational structures, external electric and magnetic fields, and spin-rotation coupling. We present rich diagrams of the system's magnetization and explain the mechanisms allowing its control on the example of molecules with spins 1/2 and 3/2. We identify the anisotropic part of the intermolecular interaction and the spin-rotation coupling as crucial for observing the system's nontrivial magnetic behavior. We propose the quench dynamics experiments to probe and reconstruct the system's molecular characteristics from observing its time evolution. We show that the strong anisotropic interaction leaves a clear mark on the system's time evolution after the quench of the interaction strength. On the other hand, the time evolution of the system's magnetization, after the electric field quench, depends significantly on the spin-rotation coupling strength. The results show an intimate coupling between the electric and magnetic properties of the system and indicate the possibility of controlling the molecular few-body magnetization with the external electric field.
In this way, we complement the previous studies on the coupling between the molecular electronic spins and external electric field in the free-space collisions \cite{Tscherbul06JCP, Tscherbul06PRL, Abrahamsson2007}. The investigated model system paves the way towards studying the controlled magnetization of ultracold molecules trapped in optical tweezers or optical lattices and their application in quantum simulation of molecular multichannel many-body Hamiltonians and quantum information storing. Recent experiments with molecules in optical tweezers~\cite{Liu18Science,Anderegg19Science} lay the grounds for the realization of the considered system. 

The plan of the paper is as follows. Section \ref{sec:model} describes the theoretical model, its experimental feasibility, and used numerical methods. Section \ref{sec:results} presents and discusses the analysis of the magnetic properties of the system and shows how the quench dynamics can unravel its underlying molecular characteristics. Section \ref{sec:summary} summarizes our paper and considers future possible applications and extensions.

%------------------------------------------------------------------------------
\section{Theoretical model and methods}
\label{sec:model}
%------------------------------------------------------------------------------

We consider two interacting distinguishable ultracold molecules bound to move along one dimension, chosen to be a $z$ axis, due to the presence of a strong transverse confinement. They are further confined in the $z$-direction by a harmonic potential of frequency $\omega$. 
The molecules are described within the rigid rotor approximation, have the same mass $m$ and spin $s$, and are in the same vibrational state. 
We approximate the interaction between molecules with the intermolecular isotropic and anisotropic contact potential.
The theoretical model is presented in detail in Ref.~\cite{DawidPRA18}, along with all necessary assumptions and approximations discussed.
The computer code allowing for reproducing the present results is available on GitHub~\cite{OurRepo}.

\subsection{Hamiltonian and basis set}

The Hamiltonian describing our system is
\begin{equation}\label{eq:tot_ham}
\hat{H}=\hat{H}_\mathrm{trap}+\hat{H}_\mathrm{mol}+\hat{H}_\mathrm{field}+\hat{H}_\mathrm{int}\,,
\end{equation}
where $\hat{H}_\mathrm{trap}$ describes the motion of molecules in a one-dimensional harmonic trap, $\hat{H}_\mathrm{mol} = \hat{H}_\mathrm{rot} + \hat{H}_\mathrm{spin-rot}$ describes the internal rotational structure of molecules and the spin-rotation coupling between the electronic spin and the rotational angular momentum of each molecule, $\hat{H}_\mathrm{field} = \hat{H}_\mathrm{Stark} + \hat{H}_\mathrm{Zeeman}$ describes the interaction of molecules with external electric and magnetic fields through the Stark and Zeeman effects, and $\hat{H}_\mathrm{int}$ describes the intermolecular interaction between molecules. Explicitly,
\begin{equation}\label{eq:partial_hams}
\begin{split}
&\hat{H}_\mathrm{trap}=\sum_{i=1}^2 \frac{{\hat{p}}^2_i}{2m}+\sum_{i=1}^2\frac{1}{2}m\omega z_i^2\,,\\
&\hat{H}_\mathrm{rot}=\sum_{i=1}^2 B\,\mathbf{\hat{j}}_i^2\,,\\
&\hat{H}_\mathrm{spin-rot}=\sum_{i=1}^2  \gamma\,\mathbf{\hat{s}}_i\cdot\mathbf{\hat{j}}_i \,,\\
&\hat{H}_\mathrm{Stark}=-\sum_{i=1}^2 \mathbf{\hat{d}}_i\cdot\mathbf{\mathcal{E}}\,,\\
&\hat{H}_\mathrm{Zeeman}= 2\mu_B\sum_{i=1}^2\mathbf{\hat{s}}_i\cdot\mathbf{\mathcal{B}}\,,
\end{split}
\end{equation}
where $B$ is the rotational constant, $\gamma$ is the spin-rotation coupling strength, $\mathbf{\hat{j}}_i$ is the $i$-th molecule's rotational angular momentum operator, $\mathbf{\hat{s}}_i$ is the $i$-th molecule's electronic spin angular momentum operator, $\mathbf{\hat{d}}_i$ is the $i$-th molecule's electric dipole moment operator, and $\mathcal{E}$ and $\mathcal{B}$ are the external electric and magnetic field strengths, respectively. For the convenience, we use units of energy and interaction strength that correspond to $\omega=m=\hbar=1$. This amounts to measuring energies in units of $\hbar\omega$, lengths in units of the harmonic oscillator characteristic length $a_\text{ho}=\sqrt{\hbar/(m\omega)}$, and interaction strengths in units of $\hbar\omega a_\text{ho}$.

The rotational constant $B$ in this study is set to $\pi \, \hbar \omega$. It corresponds to a less pronounced molecular character of the system as compared to Ref.~\cite{DawidPRA18}, where very small rotational constants were selected to enhance the impact of molecular features. Now, trap levels are more dense than rotational ones, which is a regime closer to experimental conditions (in which molecular rotational constants are usually much larger than a trap frequency). Still, we select the relatively small rotational constant to reveal an important role played by the rotational degree of freedom. With a choice of an irrational value, we additionally avoid accidental degeneracies of energy levels.

We separate the center-of-mass and relative motions in the Hamiltonian of Eq.~\eqref{eq:tot_ham} and represent the wave function of the relative motion in the following basis set:
\begin{equation}
\ket{n} \ket{J,M_J,j_1,j_2} \ket{S, M_S, s_1, s_2} \equiv \ket{\alpha} \,,
\end{equation}
which is composed of the eigenstates of the one-dimensional harmonic oscillator $\ket{n}$, eigenstates of the total rotational angular momentum operator $\mathbf{\hat{J}}$ denoted as $\ket{J, M_J, j_1, j_2}$, and eigenstates of the total electronic spin angular momentum operator $\mathbf{\hat{S}}$ denoted as $\ket{S, M_S, s_1,s_2}$.
We limit the set of $\ket{n}$ to even functions due to the trivial behavior of odd states as showed in Ref.~\cite{DawidPRA18}.
The mentioned total angular momenta are the sums of the angular momenta of individual molecules, $\mathbf{\hat{J}} = \mathbf{\hat{j}}_1 + \mathbf{\hat{j}}_2$ and $\mathbf{\hat{S}} = \mathbf{\hat{s}}_1 + \mathbf{\hat{s}}_2$.
The total angular momentum of the system is then the sum of the total rotational and spin angular momenta $\mathbf{\hat{J}}_\text{tot}=\mathbf{\hat{J}}+\mathbf{\hat{S}}$, and its projection $M_\text{tot} = M_J + M_S$ is a sum of projections of the total rotational $M_J$ and spin $M_S$  angular momenta.

The Hamiltonian describing the interaction between molecules is
\begin{equation}\label{eq:Hamint}
\hat{H}_\mathrm{int}=\hat{H}_\mathrm{iso}+\hat{H}_\mathrm{aniso}\,,
\end{equation}
where we distinguish the isotropic part $\hat{H}_\mathrm{iso}$ having the same nature as the spherically symmetric interaction between $S$-state atoms in the electronic ground state and the anisotropic part $\hat{H}_\mathrm{aniso}$ responsible for the transfer of the internal rotational angular momenta between molecules and resulting from the molecular internal structure and orientation dependence of intermolecular interactions. The more detailed discussion of various models of the anisotropic interaction is provided in Ref.~\cite{DawidPRA18}. Here, we restrict our model to the leading order of the anisotropic interaction described by the following effective Hamiltonians:
\begin{equation}
\begin{split}
\hat{H}_\mathrm{iso}&=\sum_{J, M, j_1, j_2} g_0 \delta(z_1-z_2) \hat{P}_{0} \,,\\
\hat{H}_\mathrm{aniso}&=\sum_{J, M, j_1\neq\j_1', j_2\neq\j_2'} g_{\pm 1}\delta(z_1-z_2)\hat{P}_{\pm 1}
\end{split}
\end{equation}
with
\begin{equation}
\begin{split}
\hat{P}_{0} &= \ket{J, M, j_1, j_2} \bra{J, M, j_1, j_2} \,,\\
\hat{P}_{\pm 1} &= \ket{J, M, j_1 \pm 1, j_2} \bra{J, M, j_1, j_2 \mp 1} + \text{H.c.}\,,
\end{split}
\end{equation}
where $g_0$ and $g_{\pm 1}$ are the strengths of the isotropic and anisotropic interactions, respectively, and $\delta(z)$ is the Dirac delta function imposing the contact-type interaction. 
The summation is performed over all basis set functions describing the systems' rotational degrees of freedom.

\subsection{Magnetization and quench dynamics}

In the first step, we calculate the magnetization $\langle \hat{S}_z \rangle$ of the analyzed system in the several lowest eigenstates, which is an expectation value of the $z$-component of the total electronic spin operator. In the second step, we analyze the nonequilibrium dynamics of the system after the quench.

The quench dynamics experiments may allow to probe the internal parameters of the Hamiltonian governing the analyzed system.
In a quench scenario, the system prepared initially in the chosen state $\ket{\Psi}$ (e.g. the ground state) of a Hamiltonian $\hat{H}_{\text{ini}}$, evolves unitarily in time following the sudden change (quench) of the parameters to a final Hamiltonian $\hat{H}_{\text{fin}}$~\cite{Mitra18}.
This dynamics can be expressed in terms of overlaps of the initial eigenstate $\ket{\Psi}$ of $\hat{H}_{\text{ini}}$ with the eigenstates $\ket{\tilde{\Psi}_j}$ of $\hat{H}_{\text{fin}}$:
\begin{equation}
\ket{\Psi (t)} = e^{-i \hat{H}_{\text{fin}} t} \ket{\Psi (0)} = \sum_{j} \braket{\tilde{\Psi}_j}{\Psi (0)} e^{-i E_j t} \ket{\tilde{\Psi}_j} \,.
\end{equation}
The time evolution of any observable $\hat{O}$ can be then described as:
\begin{gather*}
\bra{\Psi (t)} \hat{O} \ket{\Psi (t)} = \\
= \sum_{j, j'} \braket{\tilde{\Psi}_j}{\Psi (0)} \braket{\tilde{\Psi}_j'}{\Psi (0)} e^{-i (E_j - E_{j'}) t} \bra{\tilde{\Psi}_{j'}} \hat{O} \ket{\tilde{\Psi}_j} = \\
= \sum_j \left|\braket{\tilde{\Psi}_j}{\Psi (0)}\right|^2 \bra{\tilde{\Psi}_{j}} \hat{O} \ket{\tilde{\Psi}_j} + \\
+ 2 \sum_{j < j'} \braket{\tilde{\Psi}_j}{\Psi (0)} \braket{\tilde{\Psi}_{j'}}{\Psi (0)} \cos{[(E_j - E_{j'}) t]} \bra{\tilde{\Psi}_{j'}} \hat{O} \ket{\tilde{\Psi}_j} \,.
\end{gather*}

We choose to study the evolution of two observables: magnetization $\langle \hat{S}_z \rangle$ of the system and the cloud size $\langle \hat{r}^2 \rangle$. The formula for the cloud size is provided in ESI$\dag$. The dynamics is calculated till time $t = 10\, 000 \frac{2 \pi}{\omega}$ with a time step of $0.1 \frac{2 \pi}{\omega}$. Elongating the dynamics calculations provides no additional frequency peaks of amplitudes larger than $10^{-4}$ in the corresponding discrete Fourier transforms (DFT). We perform the Fourier transform of the observables' evolution using SciPy package~\cite{SciPy}.

\subsection{Convergence with the basis set size}

The systems' eigenstates are calculated using the exact diagonalization method with the basis set composed of quantum harmonic oscillator eigenfunctions up to $n_{\text{max}} = 20$, and quantum rigid rotor eigenfunctions up to $j_{\text{max}} = 4$. 
The slow convergence with $n_{\text{max}}$ is asymptotically proportional to $\frac{1}{\sqrt{n_{\text{max}}}}$~\cite{Grining15}, however in the analyzed system $n_{\text{max}} = 20$ provides a satisfying convergence for the lowest-energy states, which are the main focus of this paper.

The convergence of the cloud-size time-evolution calculations with $n_{\text{max}}$, however, is problematic. On the one hand, the mean value of the cloud size, $\bra{n} \hat{r}^2 \ket{n'}$, calculated for two harmonic oscillator eigenfunctions, increases rapidly with the harmonic oscillator levels, $n$, becoming divergent for large $n$ (see ESI$\dag$). In the quench dynamics, this divergence is faster than the decrease of the overlap between the ground state and the highly excited states leading to the divergence of the cloud-size excitation. However, this nonphysical behavior can be neglected by restricting the basis sets size, knowing that all realistic traps have a finite size, and all realistic quenches have a finite time. On the other hand, the highest-energy eigenstates in a finite basis set are not converged~\cite{Busch98,Grining15} and a nonphysically larger occupation of the highest-energy eigenstate can be observed. We solve this problem by neglecting unconverged part of spectrum from the quench dynamics calculations, namely by removing eigenfunctions $\ket{\tilde{\Psi}_j}$ of $\hat{H}_{\text{fin}}$ with the contribution from any basis state with $\ket{n_{\text{max}}}$ larger than 10\%. The removed part of eigenfunctions is a few percent of the whole spectrum.

According to our previous convergence analysis~\cite{DawidMSc17}, selected $j_{\text{max}}$ already enables convergence for the significant part of the spectrum as well as good convergence of the quench dynamics.

\subsection{Experimental feasibility}\label{ssec:exp_feas}

The smaller the ratio of the molecular rotational constants to the harmonic trap frequency, $B/\omega$, the more pronounced the analyzed system's molecular character~\cite{DawidPRA18}. However, typical ultracold molecular experimental set-ups~\cite{Ni08, Molony14, Park15} use three-dimensional trap frequencies ranging from around 1 kHz to at most 1 MHz~\cite{Bolda02}, while studied ground-state molecules have rotational constants reaching hundreds of MHz or more~\cite{DeiglmayrJCP08}. This combination amounts to the ratio of $B/\omega \gg 1$ and results in the rotational levels separated by many harmonic trap states. In such a scenario, the molecular features related to the molecular rotational structure are less important. 

\begin{figure}[t]
\begin{center}
\includegraphics[width=\columnwidth]{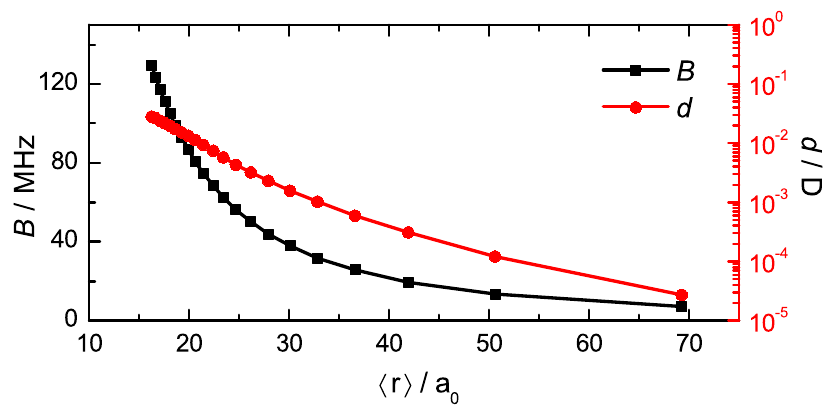}
\end{center}
\caption{The rotational constant, $B$, and estimated electric dipole moment, $d$, of the exemplary ${}^{87}$Rb${}^{133}$Cs molecule in the lowest triplet state, as functions of the mean distance between atoms $\langle r \rangle$, corresponding to different vibrational levels. Similar behavior may be expected for other weakly-bound molecules.}
\label{fig:properties}
\end{figure}

In this work, we analyze the regime of $B/\omega \approx 3$, which can be reached with tight traps (e.g.,~a nanoplasmonic one~\cite{ThompsonScience13}) and weakly-bound molecules (e.g., Feshbach molecules~\cite{Chin10}). Feshbach molecules have rotational constants up to few MHz, but their electric dipole moments may be vanishingly small, as they scale asymptotically as $R^{-7}$ with the internuclear separation $R$~\cite{Lahaye09}. Figure~\ref{fig:properties} presents the dependence of the rotational constant and permanent electric dipole moment on the mean distance between atoms in the ${}^{87}$Rb${}^{133}$Cs molecule in the lowest electronic state with non-zero spin, i.e., $a^3 \Sigma$. The choice of this species is just exemplary as we expect other classes of weakly-bound molecules to have similar characteristics. One of the highest vibrational levels of $a^3 \Sigma\,{}^{87}$Rb${}^{133}$Cs, $v=39$, corresponds to the mean distance between atoms equal to 51 bohr. We estimate the corresponding electric dipole moment of $10^{-4}\,$D and rotational constant of $13.5\,$MHz. To reach the $B/\omega$ ratio of the order of the selected one, the trap frequency needs to be at least around $0.5\,$MHz. Note that the upper limit for $\omega$ is set also by the size of two molecules, which increases for weakly-bound states. The choice of the trap frequency determines the time scale used within this work. For example, the time evolution plotted in Figs.~\ref{fig:r2quench} and~\ref{fig:Mgquench} takes $200 \,(2 \pi / \, \omega) \approx 2.5 \,$ms for $\omega=0.5\,$MHz.

Within our work, we also analyze the time evolution of the cloud size after the quench of the intermolecular interaction. The sudden change of the intermolecular interaction strength can be achieved via the change of the magnetic field strength and related Feshbach resonances, the change of the molecular vibrational state, or the change of the trapping frequency. However, the selective quench of intermolecular interaction is impossible to achieve, as, in reality, all molecular characteristics are intimately connected that leads to emergent behavior and challenging description. For example, changing the vibrational state impacts the interaction but also the polarization of molecules that modifies their response to external fields. The quench of the trapping frequency should correspond to the most uncorrelated change of the intermolecular interaction~\cite{GarciaMarch16}. Regarding the observables whose time evolution we study, the internal state of molecules can be probed with quantum gas microscopy \cite{Bakr09Nature, Sherson10Nature, Boll16Science}, and the cloud size can be measured via destructive time-of-flight experiments as realized for ultracold atoms \cite{Bucker09NJP, Fuhrmanek10NJP, Bergschneider18PRA}.

\begin{figure*}[tb!]
\begin{center}
\includegraphics[width=\textwidth]{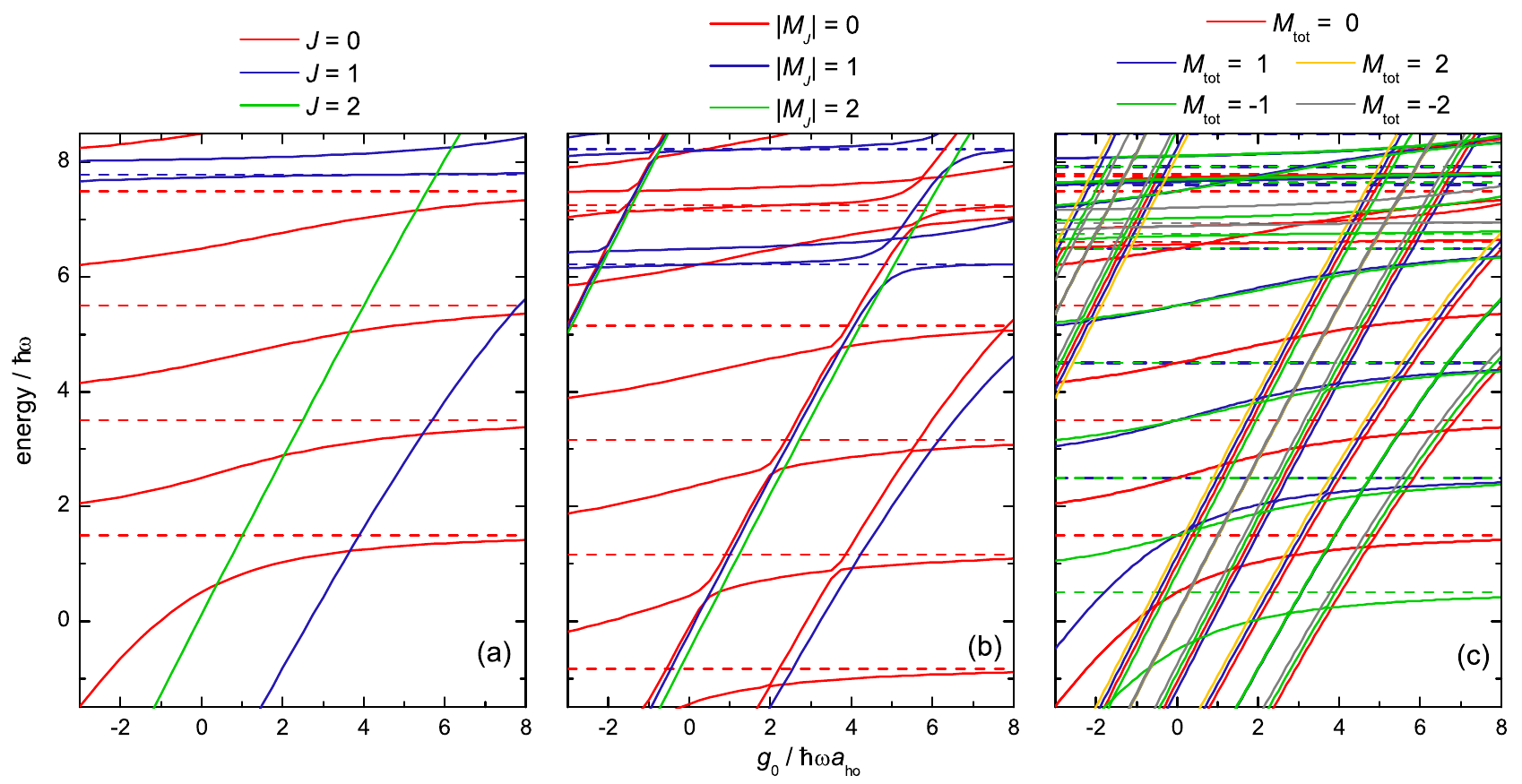}
\end{center}
\caption{Energy spectra of the relative motion for two interacting molecules with the spin $s=1/2$ and rotational constant $B= \pi \,\hbar\omega$ in a one-dimensional harmonic trap as a function of the isotropic interaction strength $g_0$ with the anisotropic interaction strength $g_{\pm 1} = 10$ and (a) no external fields, (b) electric field strength $d \mathcal{E}=5\,\hbar\omega$, and (c) the magnetic field strength $\mathcal{B}=0.5\,\hbar\omega/\mu_B$ and spin-rotation coupling constant $\gamma=0.3\,\hbar\omega$. Solid and dashed lines are for states of even and odd spatial symmetries, respectively.}
\label{fig:spectrum}
\end{figure*}

Non-reactive trapped alkali dimers in the lowest rovibrational states have been recently shown to form four-atom complexes that are long-lived in the dark but are prone to decay under the trapping field that results in losses \cite{Hu19Science,Liu20NatPhys, Gregory20PRL}. The lifetime of such complexes is proportional to the density of states at the collision threshold. The density of states of weakly-bound molecules, compared to deeply-bound, can be up to eight orders of magnitude smaller \cite{Christianen19PRA}. Therefore, the weakly-bound molecules may be less prone to such losses.
Additionally, vibrationally-excited molecules may undergo reactive collisions and other decoherence or loss processes may occur, whose detailed characterization is, however, out of the scope of the present study. Search for the molecules which exhibit low losses in a trap is a challenging and very impactful task. They probably should be lighter and have a less dense spectrum of electronic states such as AlF \cite{Truppe19PRA}. Regarding the non-zero or large electronic spin, it may be realized with alkaline-earth-metal fluoride molecules in the doublet $^2\Sigma^+$ electronic state~\cite{ShumanNature10}, alkali-metal molecules in the triplet $^3\Sigma^+$ electronic state~\cite{TomzaPRA13,RvachovPRL17}, or molecules containing highly-magnetic atom~\cite{ZarembaPRA18,FryePRX20}, respectively. Thus, the considered system may potentially be realized in state-of-the-art experiments on ultracold molecules trapped in optical tweezers~\cite{Liu18Science,Anderegg19Science}. However, exact experimental conditions have to be yet carefully researched and designed.

%------------------------------------------------------------------------------
\section{Results}
\label{sec:results}
%------------------------------------------------------------------------------

\begin{figure*}[tb!]
\begin{center}
\includegraphics[width=\textwidth]{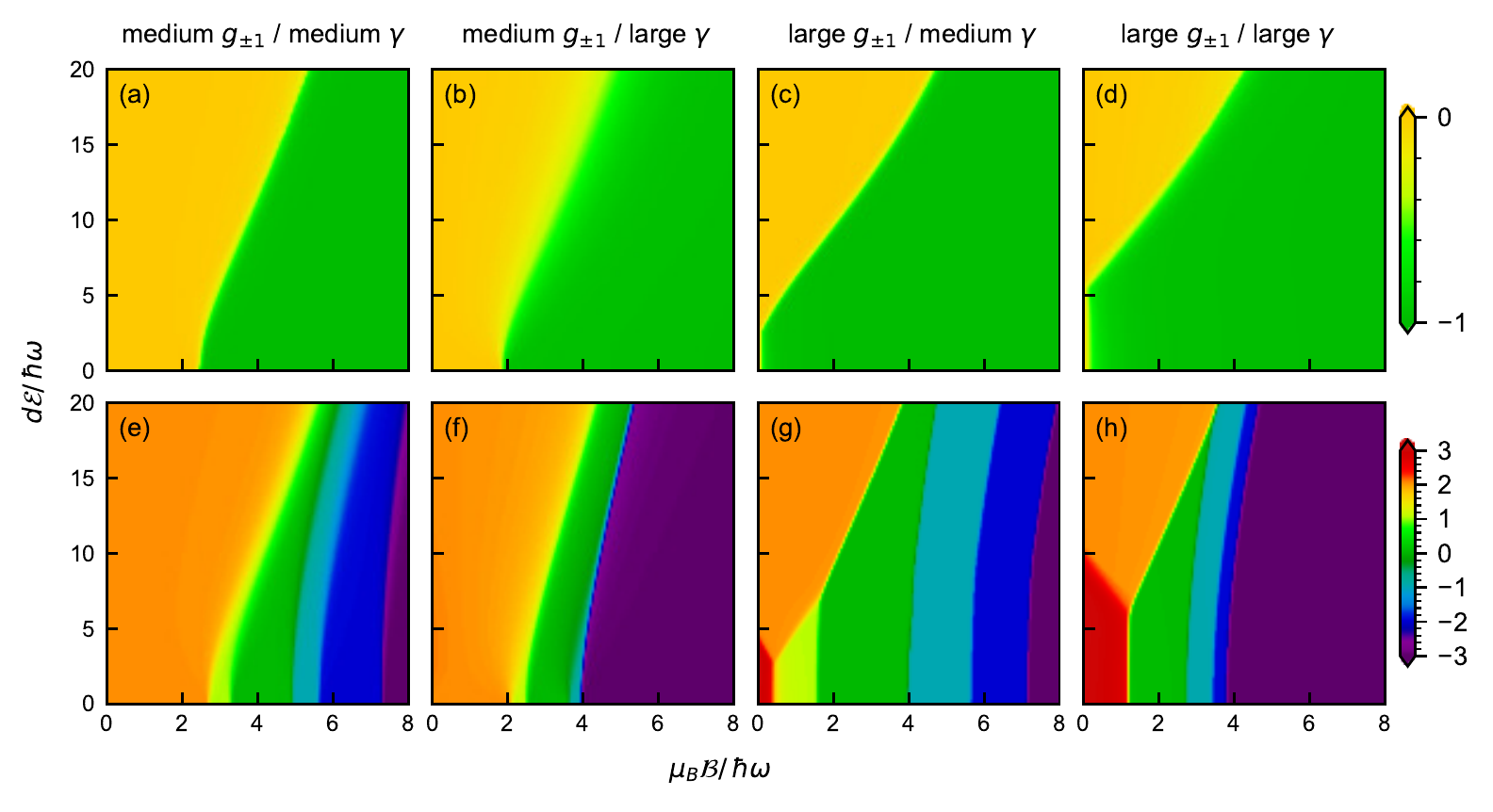}
\end{center}
\caption{Ground-state magnetization diagrams as functions of the electric $d\mathcal{E}$ and magnetic $\mu_{\text{B}}\mathcal{B}$ field strengths for two interacting molecules with the rotational constants $B= \pi \,\hbar\omega$ in a one-dimensional harmonic trap for medium isotropic interaction strength $g_0 = 4$. Panels (a)-(d) and (e)-(h) present results for spins $s=1/2$ ($M_{\text{tot}}=0$) and $3/2$ ($M_{\text{tot}}=2$), respectively.
The first two and last two columns show results for medium $g_{\pm 1} = 4$ and large $g_{\pm 1} = 10$ anisotropic interaction strengths, respectively. Moreover, the first and third columns present results for a medium spin-rotation coupling $\gamma = 1 \,\hbar\omega$, and the second and the fourth columns - for strong spin-rotation coupling $\gamma = 3 \,\hbar\omega$, respectively.}
\label{fig:phase_diagrams}
\end{figure*}

We begin by studying the magnetic properties of two interacting ultracold molecules in a one-dimensional harmonic trap. We focus on the system's magnetization. We analyze how it depends on the intermolecular interaction between molecules and the coupling between the electronic spins and the molecular rotational momenta. We present how the system's magnetization can be controlled and how this control depends on the molecular properties. Next, we study the quench dynamics designed to extract the strengths of the spin-rotation coupling as well as the isotropic and anisotropic interaction strengths between the molecules.

\subsection{Magnetic properties and its control}
\label{ssec:magnetic_props}

Mechanisms that allow control over the system's magnetization can be observed from the energy spectra. Figure~\ref{fig:spectrum} shows calculated energy spectra as functions of the isotropic interaction strength $g_0$. We select a set of internal and external Hamiltonian parameters to present the interplay between the system's magnetic properties and the external fields.

Figure~\ref{fig:spectrum}(a) presents the energy spectrum of two molecules in a one-dimensional harmonic trap with a strongly anisotropic intermolecular interaction $g_{\pm 1} = 10$ without external fields or spin-rotation coupling. The comparison with the neighboring panel (b) shows the impact of the external electric field on the system. The electric field moves the energy levels by a Stark shift and removes their degeneracy with respect to the total rotational angular momentum, $J$. The electric field splits states with $J=1$ and $J=2$ into two and three states, respectively, according to the different number of possible projections of total rotational angular momentum, $ M_J $. Shifted states then often anticross due to the coupling between different total rotational momentum states.

A comparison of panels (a) and (c) in Fig.~\ref{fig:spectrum} shows the impact of the medium magnetic field on the spectrum of two molecules with a small spin-rotation coupling ($\gamma = 0.3 \, \hbar \omega$). The only conserved quantum number is $M_{\text{tot}}$, i.e., the sum of projections of total rotational, $M_J$, and spin, $M_S$, angular momenta. States split accordingly to the Zeeman shift. States with the positive projection of the total spin angular momentum $M_S$ are high-field seekers, while energies of states with negative $M_S$ decrease with the magnetic field strength. 
Shifted states then often anticross due to a non-zero spin-rotation coupling, which mixes states with different $J, S, M_J$, and $M_S$ (see the corresponding Hamiltonian elements in ESI$\dag$). The impact of the spin-rotation coupling on the system increases with the absolute values of states' total rotational and spin angular momenta' projections, $M_J$ and $M_S$. The larger $|M_J|$ and $|M_S|$, the more numerous are possible combinations of the projections of individual rotational and spin angular momenta, $m_i, m_{s_i}$, which are mixed by the spin-rotation coupling.

Figure~\ref{fig:phase_diagrams} presents rich magnetization diagrams of the studied system in the ground state as functions of the magnetic and electric fields. The upper row shows results for two molecules with electronic spins 1/2 and $M_{\text{tot}}=0$, while the bottom one - with spins 3/2 and $M_{\text{tot}}=2$. The primary observation is that the number of possible magnetization values of the ground state is limited by the total electronic spin momenta of molecules and the selected $M_{\text{tot}}$ value. This restriction results from the definition of $M_{\text{tot}} = M_J + M_S$, but also because low-energy states are characterized by small values of rotational angular momenta, $j_1$ and $j_2$, resulting in $M_J$ values, which are close to zero.

The main reason for all magnetization changes in Fig.~\ref{fig:phase_diagrams} is the interplay between the Zeeman and Stark effects. 
The magnetic field linearly brings down energies of the states with negative $ M_S $, with speed depending on the $ M_S $ value. 
The Stark effect lowers the ground state's energy, composed mostly of the basis state with $J=M=j_1=j_2=0$, faster than the lowest state with negative magnetization, composed mostly of the basis state with $J=M=1$. Therefore, larger external electric field strengths effectively force larger magnetic field strengths for the magnetization change to happen, when the lowest states exchange their order. Such underlying interplay is visible in all panels of Fig.~\ref{fig:phase_diagrams}.

\begin{figure}[t!]
\begin{center}
\includegraphics[width=\columnwidth]{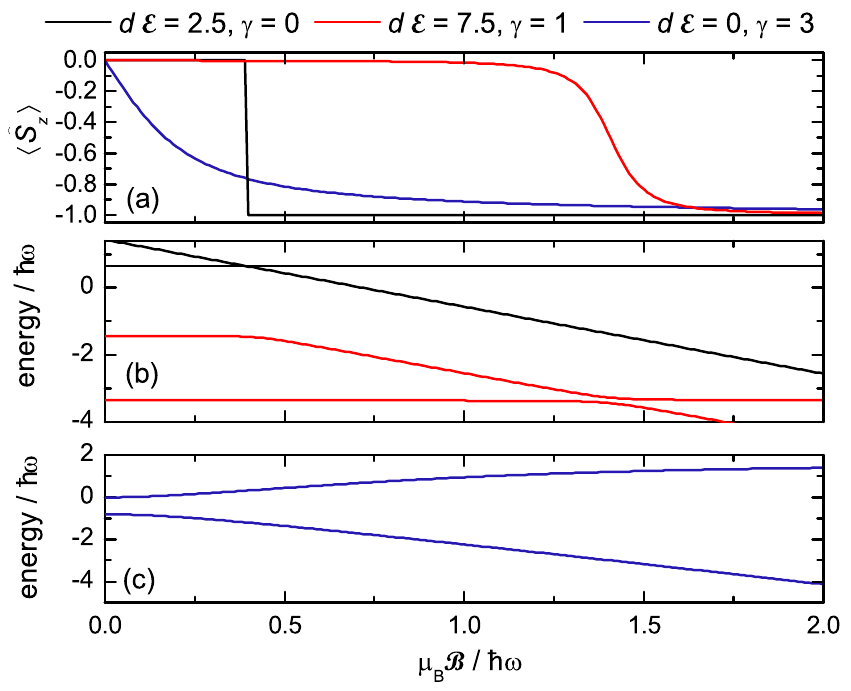}
\end{center}
\caption{(a) Ground-state magnetization of two interacting molecules with the rotational constants $B=\pi\,\hbar\omega$, spins $s=1/2$, and $M_{\text{tot}}=0$ in a one-dimensional harmonic trap with medium isotropic and strong anisotropic intermolecular interaction strengths ($g_0 = 4$ and $g_{\pm 1} = 10$) as a function of the magnetic field strength $\mu_{\text{B}} \mathcal{B}$ for different electric field strengths $d \mathcal{E}$ and spin-rotation coupling constants~$\gamma$. (b)-(c)~Energies of the analyzed ground states and coupled lowest-energy excited states as functions of the magnetic field strength $\mu_{\text{B}} \mathcal{B}$. 
Color code indicates the strengths of the electric field $d \mathcal{E}$ and the spin-rotation coupling $\gamma$.}
\label{fig:magnetisation_1/2}
\end{figure}

\subsubsection{Magnetization for spins 1/2}
\label{sssec:Mg_spins12}

Panels (a)-(d) of Fig.~\ref{fig:phase_diagrams} present the ground-state magnetization diagrams for the system composed of two interacting molecules with spins 1/2 and $M_{\text{tot}} = 0$. The choice of $M_{\text{tot}}$ limits the number of possible $\langle \hat{S}_z \rangle$ values. In this case, two states are mainly responsible for the magnetization change, namely the ground state dominated by the $J=M_J=j_1=j_2=0$ and $M_S = 0$ basis state and the excited state dominated by the $J=M_J=1$ and $M_S = -1$ basis state.

Panels (a)-(b) of Fig.~\ref{fig:phase_diagrams} show the magnetization diagrams for medium strength of the anisotropic interaction ($g_{\pm 1} = 4$). In the absence of the spin-rotation coupling, the magnetization change results simply from the Zeeman and Stark effects' interplay. 
Figure ~\ref{fig:magnetisation_1/2}(b) depicts in black the energies of states taking part in such a change. The black line in the panel (a) of the same figure presents the resulting $\langle \hat{S}_z \rangle$ of the ground state. The electric field can control the magnetic field's strength at which the change takes place.

The sharply crossing states are not coupled either by the electric field (which conserves $M_J$) or the spin-rotation coupling (which conserves $j_1$ and $j_2$). However, if both couplings are present, the intermediate state dominated by the $J=1, M_J=0$ basis state, provides the second-order coupling between the discussed states, which results in their anticrossing visible in red in Fig.~\ref{fig:magnetisation_1/2}(a)-(b). The repulsion between states grows with the spin-rotation coupling strength, as seen when comparing the magnetization diagrams in panels (a) and (b) of Fig.~\ref{fig:phase_diagrams}. The spin-rotation coupling also lowers the energy of states with the largest absolute values of $M_J$ and $M_S$, allowing for the magnetization change in the weaker magnetic field.

The importance of the intermolecular interaction anisotropy is visible when comparing the first two columns ($g_{\pm 1} = 4$) with the last two ones ($g_{\pm 1} = 10$) of Fig.~\ref{fig:phase_diagrams}, i.e., panels (a)-(b) with (c)-(d). Firstly, the systems with the larger anisotropic interaction strength require smaller external fields for the ground-state magnetization change related to the crossing of the lowest-energy states. For electric field strengths larger than a few $\hbar \omega$, the $\langle \hat{S}_z \rangle$ change takes place within the same two mechanisms described above. States with different magnetization either cross in the absence of the spin-rotation coupling or anticross when both the electric field and spin-rotation coupling are present. However, the intermolecular interaction's large anisotropy allows an additional mechanism for small electric field strengths. It brings down the states with higher total rotational momenta, including $J = 1$ (as seen in Fig.~\ref{fig:spectrum} (b)), and non-zero $M_J$ and $M_S$. So when the magnetic field, through the Zeeman effect, lifts the degeneracy of $M_S$, the two states strongly repel each other thanks to the spin-rotation coupling, as seen in blue in Fig.~\ref{fig:magnetisation_1/2}(a),(c).
The larger the spin-rotation coupling strength, the larger electric field strength is needed to push down the state with $J=0$ and reproduce the mechanisms described for smaller $g_{\pm 1}$.

\subsubsection{Magnetization for spins 3/2}
\label{sssec:Mg_spins32}

\begin{figure}[t!]
\begin{center}
\includegraphics[width=\columnwidth]{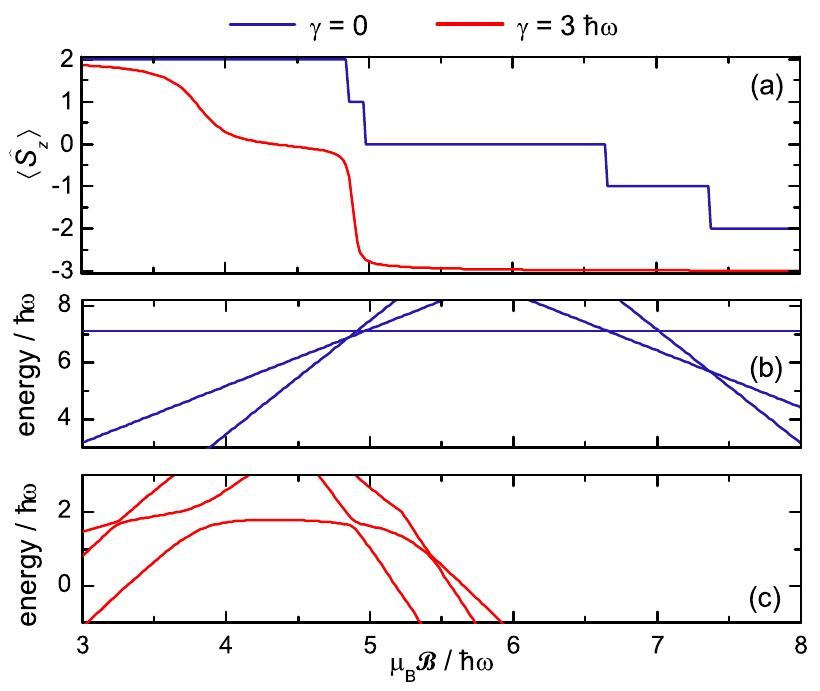}
\end{center}
\caption{(a) Ground-state magnetization of two interacting molecules with the rotational constants $B=\pi\,\hbar\omega$, spins $s=3/2$, and $M_{\text{tot}}=2$ in a one-dimensional harmonic trap with medium isotropic and anisotropic intermolecular interaction strengths ($g_0 =  g_{\pm 1} = 4$) as a function of the magnetic field strength $\mu_{\text{B}} \mathcal{B}$ for the electric field strength $d \mathcal{E} = 15 \, \hbar \omega$ and different spin-rotation coupling constants $\gamma$. (b)-(c)~Energies of the analyzed ground states and coupled lowest-energy excited states as functions of the magnetic field strength $\mu_{\text{B}} \mathcal{B}$. Color code indicates the spin-rotation coupling constants $\gamma$.}
\label{fig:magnetisation_3/2}
\end{figure}

Panels (e)-(h) of Fig.~\ref{fig:phase_diagrams} show the ground-state magnetization diagrams of two interacting molecules with spins $3/2$ and $M_{\text{tot}} = 2$. This choice results in a larger number of possible $\langle \hat{S}_z \rangle$ values, ranging from -3 to 3.
Also, the states taking part in the magnetization changes have much higher rotational angular momenta than ones in the previous section.

In the case of the medium anisotropic interaction strength, the choice of $M_{\text{tot}} = 2$ in the absence of external fields results in a ground state with $M_J = 0, M_S = 2$. In the absence of the electric field or the spin-rotation coupling, the lowest states cross each other due to the Zeeman effect, resulting in abrupt ground-state magnetization changes. Blue lines in Fig.~\ref{fig:magnetisation_3/2}(a)-(b) present an example of the magnetization and energies of such states as a function of the magnetic field strength.

When the spin-rotation coupling is present, the spectra become dense and exhibit multiple anticrossings, due to mixing of $ J $, $ S $, and projections of individual molecular rotational and spin angular momenta $m_1, m_2, m_{s_1}, m_{s_2}$. The anticrossing strength depends on two factors. The strongest anticrossings occur between states belonging to the same harmonic level, as both the spin-rotation coupling and electric field conserve $n$.
Another factor is the anticrossing states' composition of individual rotational angular momenta, $j_1$ and $j_2$.
The larger difference between them, the smaller is the coupling induced by the electric field.
This is why the anticrossing strength decreases with the difference of states' magnetization, as presented in Fig.~\ref{fig:magnetisation_3/2}, where states depicted in red change the magnetization from 2 to 0 and 0 to -3.
Due to conserved $M_{\text{tot}}$, the change of $M_S$ is compensated by the increase of $M_J$.
Larger $M_J$ forces higher rotational momenta, $j_1$ and $j_2$.
Therefore, the larger magnetization change, the larger difference between the states' rotational momenta, and the smaller coupling between anticrossing states.

The spin-rotation coupling's stronger impact on states with large absolute values of $M_S$ and $M_J$ is more prominent for spins 3/2 than for spins 1/2.
A comparison between panels (e) and (f) as well as (g) and (h) of Fig.~\ref{fig:phase_diagrams} shows that the larger spin-rotation coupling can not only bring the magnetization change to lower magnetic field strengths, but also effectively limits the number of accessible magnetization values, as in the case of panels (f) and (h).

The same dependencies determine the magnetization diagrams for the systems with large anisotropy of the intermolecular interaction ($g_{\pm 1}=10$), as seen in panels (g) and (h) of Fig.~\ref{fig:phase_diagrams}.
The main difference comes from an additional $\langle \hat{S}_z \rangle$ value accessible for weaker electric fields due to bringing down the state with higher rotational angular momentum with $\langle \hat{S}_z \rangle = 3$ by the anisotropy of the intermolecular interaction.

%------------------------------------------------------------------------------
\subsection{Quench dynamics}
\label{ssec:quenches}
%------------------------------------------------------------------------------

The analyzed system's magnetic properties depend strongly on the anisotropic part of the intermolecular interaction and the spin-rotation coupling. These molecular properties are challenging to calculate or measure. However, they can be extracted through the analysis of the quench dynamics of observables that they influence. 

Therefore, first, we study the nonequilibrium dynamics of the cloud size, $\langle \hat{r}^2 \rangle$, after the quench of the intermolecular interaction which can be achieved via the change of the trapping frequency~\cite{GarciaMarch16} (see the discussion in section \ref{ssec:exp_feas}). We aim at identifying dynamical signatures of isotropic and anisotropic part of the interaction between the molecules. Next, we analyze the time evolution of the magnetization, $\langle \hat{S}_z \rangle$, after the quench of the external electric or magnetic field. It provides insight into the strength of the spin-rotation coupling present in the molecular system.

To reconstruct couplings governing the dynamics, we perform the discrete Fourier transform (DFT) of the corresponding time evolution. The resulting function indicates the frequencies dictating the time evolution. These frequencies can then be transformed into energy differences between states whose coupling causes the system's nontrivial dynamics. The strength of the coupling is related to the peak's amplitude at the corresponding frequency.

\subsubsection{The isotropic and anisotropic intermolecular interactions}
\label{sssec:quench_r2}

\begin{figure}[t]
\begin{center}
\includegraphics[width=\columnwidth]{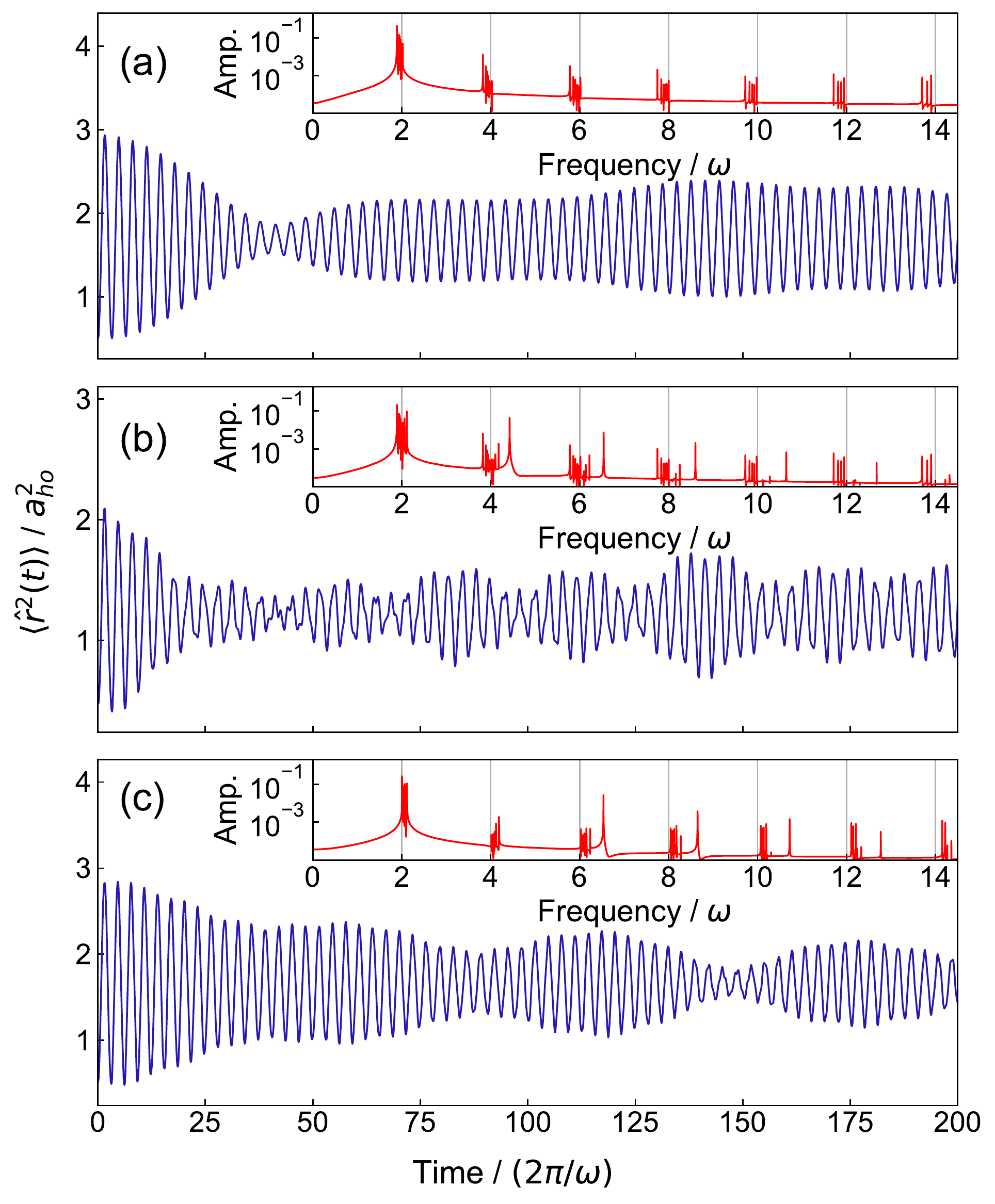}
\end{center}
\caption{The time evolution of the cloud size, $\langle \hat{r}^2 \rangle$, after the quench of the intermolecular interaction strength between two interacting molecules without spin and with the rotational constants $B=\pi\hbar\omega$ in a one-dimensional harmonic trap with the total rotational angular momentum $J=1$. The isotropic and anisotropic interaction is quenched from zero $g_0 = g_{\pm 1} =0 $ to (a) $g_0 = 4$, $g_{\pm 1} = 0$, (b) $g_0 = 0$, $g_{\pm 1} = 4$, and (c) $g_0 = 4$, $g_{\pm 1} = 10$. Insets present the discrete Fourier transforms of the studied time evolutions.}
\label{fig:r2quench}
\end{figure}

Figure \ref{fig:r2quench} presents the nonequilibrium dynamics of the analyzed molecular system with total rotational angular momentum, $J=1$, after the quench of the interaction, starting from the noninteracting case, with the initial state $\ket{\Psi_0} = \ket{n=0, \, J=1, \, j_1=0, \, j_2=1}$.

Panel (a) shows the time evolution of the cloud size after quenching the isotropic part of the intermolecular interaction from zero to medium strength $g_0 = 4$, which couples states composed of different harmonic trap levels. The inset of panel (a) presents the DFT of the studied time evolution, which can be used to unravel couplings between states governing the quench dynamics. The molecular system's dynamics in these conditions is almost identical to two ultracold atoms in a one-dimensional harmonic trap, even though the rotational structure is present, and $ J $ is nonzero. The reason is the lack of the anisotropic part of the intermolecular interaction and the absence of couplings between the rotational states. As a result, the multiple peaks visible in the DFT correspond to the couplings with different harmonic states only, which are almost exactly evenly separated by $2\, \omega$ (we ignore odd states in this work). The divergence from the single ladder of frequencies comes from the slightly different influence that the isotropic interaction has on the system's ground state, as compared to excited ones, as known for both atomic \cite{Busch98} and molecular \cite{DawidPRA18} cases. The additional ladder of frequencies slightly below $2 \, \omega$ comes from the couplings of higher-energy harmonic states to the ground state. If the isotropic intermolecular interaction is quenched to negative strengths, the multiple peaks would become more detached. In the case of the quench to the strongly repulsive interaction, entering the Tonks-Girardeau regime, the DFT would show a single ladder of peaks evenly separated by $2 \, \omega$. The analysis above is independent of the total rotational angular momentum value, $ J $ as long as the anisotropic part of the interaction is zero.

Panel (b) of Fig.~\ref{fig:r2quench} presents the dynamics after the quench of the anisotropic part of the intermolecular interaction from zero to medium strength $g_{\pm 1} = 4$, keeping the isotropic part equal to zero. The anisotropic interaction couples not only different harmonic trap states but also rotational ones, preserving $J$. It impacts the spectrum in two ways. Firstly, as showed in Ref.~\cite{DawidPRA18}, the anisotropic interaction splits each harmonic trap state with $J=1$ into two states, the antisymmetric and symmetric one. The splitting depends slightly on the harmonic level and is largely similar across the spectrum, except for the lowest-energy state. The antisymmetric ground state, resulting from the splitting of the lowest-energy harmonic state, is brought down rapidly by the anisotropic part of the interaction. Both effects can be seen in the DFT of the time evolution of the cloud size in Fig.~\ref{fig:r2quench}(b) as well as in the system's spectrum presented in Fig.~\ref{fig:spectrum}(a). The described splitting results in two close ladders of excited symmetric and antisymmetric eigenstates with frequencies close to $2\, \omega$. The ladder of frequencies separated by $\approx 1.9\, \omega$ comes from the couplings between the symmetric states, while the neighboring ladder starting from $\approx 2.1\, \omega$ results from the couplings between the antisymmetric states. The couplings with the separated ground state cause an additional ladder starting for a frequency equal to the energy difference between the ground energy and the nearest excited antisymmetric state (here around $4.4 \, \omega$). The ground state's sensitivity to the anisotropic interaction strength allows for using this frequency as a quite precise signature of this molecular property. Invisible for the quench dynamics are states with $j_1 = j_2$, as nothing couples them to the system's initial state. If the initial state, $\ket{\Psi_0}$, is antisymmetric, instead of composed solely by $\ket{n=0, \, J=1, \, j_1=0, \, j_2=1}$, the only couplings governing the system are the ones between antisymmetric states. The time evolution of $\langle \hat{r}^2 \rangle$ is then simpler, and the DFT contains no splittings of the main frequency ladder. If $\ket{\Psi_0}$ is chosen to be symmetric, the couplings to the antisymmetric ground state disappear, rendering this quench scenario less sensitive to the anisotropic interaction strength.

Panel (c) of Fig.~\ref{fig:r2quench} shows the time evolution of the cloud size after quenching both parts of the intermolecular interaction, i.e., the isotropic part from zero to medium strength $g_0 = 4$ and anisotropic part from 0 to large strength $g_{\pm 1} = 10$. In this case, the DFT shows a frequency ladder similar to the one observed in panel (a) and the additional ladder starting from the frequency around $6.5\, \omega$ analogous to the panel (b). However, the result is not a simple sum of two interaction parts' effects, but it rather comes from the competition between them. Firstly, the ground state, detached from the rest of the spectrum by the anisotropic part of the interaction, is pushed back up by the isotropic part, as seen in Fig.~\ref{fig:spectrum}(a). For a molecular system with the dominantly isotropic interactions or with an anisotropic part equal to isotropic, the separated ground state is gone and the additional ladder in the corresponding DFTs. On the other hand, the larger frequency at which the additional ladder starts, the larger is the dominance of the anisotropy over the isotropy of the intermolecular interaction. The splittings between ladders starting around $2 \, \omega$ are another signature of the intermolecular interaction. They get negligible small, resulting in a single ladder, in two cases. The first case is entering the Tonks-Girardeau regime with a very strong isotropic part of the interaction. Second, when the anisotropic part of the interaction's strength is equal to the isotropic one. This regime corresponds to the metastable gas-like super-Tonks states \cite{DawidPRA18, HallerScience09, AstrakharchikPRL05, TempfliNJP08}, being a molecular equivalent of the Tonks-Girardeau regime.

The results showed in Fig.~\ref{fig:r2quench} and discussed above concern the system with the total rotational angular momentum $J=1$. They present the possibility of extracting the relative strength of the anisotropic part of the intermolecular interaction, $g_{\pm 1}$, compared to the isotropic part, $g_0$. On the other hand, the quench analysis of the system with a zero total rotational angular momentum allows determining $g_0$. The reason is the lack of the dependence of eigenstates with $J=0$ on the anisotropic part of the intermolecular interaction. Therefore, to extract the full information about intermolecular interactions, one should start with the interaction quench performed in the system with $J=0$, interpreted as in Fig.~\ref{fig:r2quench}(a), followed by the investigation of the quench dynamics of the system with $J=1$ or higher.

\subsubsection{The spin-rotation coupling}
\label{sssec:quench_Mg}

\begin{figure}[t]
\begin{center}
\includegraphics[width=\columnwidth]{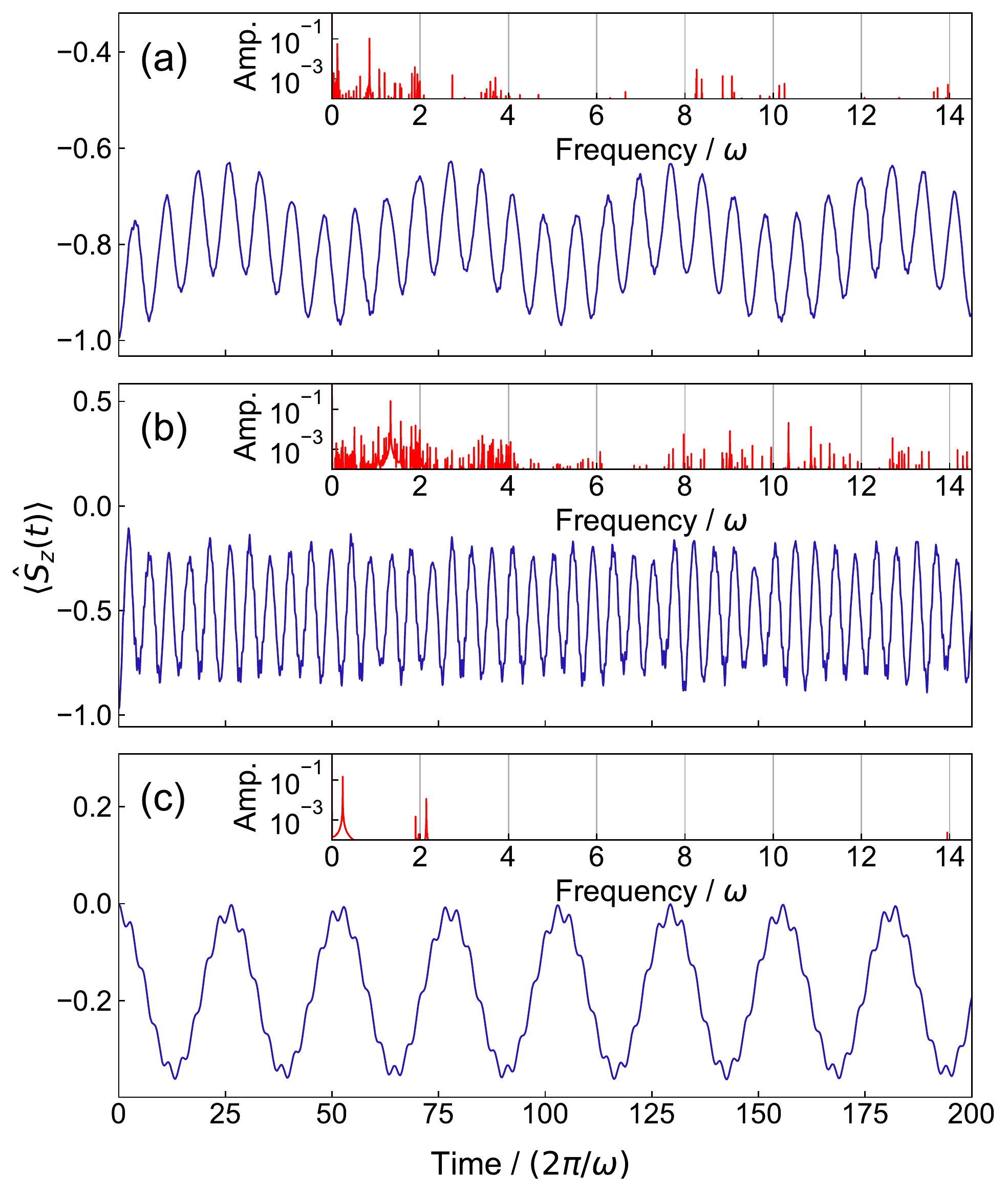}
\end{center}
\caption{The time evolution of the magnetization, $\langle \hat{S}_z \rangle$, of the system of two interacting molecules with the spins 1/2 and rotational constants $B=\pi\hbar\omega$ in a one-dimensional harmonic trap, described by the medium isotropic and anisotropic interaction strengths, $g_0 = g_{\pm 1} = 4$, and with zero projection of the total angular momentum, $M_{\text{tot}} = 0$, after the quench of (a) the electric field from 0 to $d \mathcal{E} = 7.5 \, \hbar \omega$ for a system with spin-rotation coupling $\gamma = 1 \, \hbar \omega$ and with the constant magnetic field, $\mu_B \mathcal{B} = 3 \, \hbar \omega$, (b) the electric field from 0 to $d \mathcal{E} = 7.5 \, \hbar \omega$ for a system with $\gamma = 3 \, \hbar \omega$ and constant magnetic field, $\mu_B \mathcal{B} = 3 \, \hbar \omega$, and (c) the magnetic field from 0 to $\mu_B \mathcal{B} = 4 \, \hbar \omega$ with $\gamma = 1 \, \hbar \omega$ and the constant electric field $d \mathcal{E} = 5 \, \hbar \omega$. Insets present the discrete Fourier transforms of the studied time evolutions.}
\label{fig:Mgquench}
\end{figure}

Figure \ref{fig:Mgquench} presents the nonequilibrium dynamics of the magnetization, $\langle \hat{S}_z \rangle$, of the analyzed molecular system with molecular electric spins 1/2 with zero projection of the total angular momentum, $M_{\text{tot}} = 0$, after the quench of external electric or magnetic field with different spin-rotation coupling strengths, $\gamma$. To see a nontrivial time evolution of $\langle \hat{S}_z \rangle$, i.e., its value changing in time with a significant amplitude, the initial state, $\ket{\Psi_0}$, must be coupled to a subset of the final eigenfunctions, $\ket{\tilde{\Psi}_j}$, with $\bra{\tilde{\Psi}_j} \hat{S}_z \ket{\tilde{\Psi}_{j'}}\neq 0$. These conditions are not met in the system without the spin-rotation coupling, as it is the only part of the Hamiltonian, which mixes states with different projections of the total spin angular momentum, $M_S$. Therefore, in general, the smaller $\gamma$, the smaller amplitude of $\langle \hat{S}_z (t)\rangle$ after a quench of any external field.
However, the presence of the spin-rotation coupling is not a sufficient condition for the nontrivial dynamics. The quench of the external fields must also be performed in the vicinity of the magnetization change, discussed in Sec.~\ref{ssec:magnetic_props}, otherwise the value of $\bra{\tilde{\Psi}_j} \hat{S}_z \ket{\tilde{\Psi}_{j'}}$ becomes negligibly small.

Panels (a) and (b) of Fig.~\ref{fig:Mgquench} present the time evolution of the magnetization, $\langle \hat{S}_z \rangle$, for the system under the constant impact of the external magnetic field of $3\, \hbar \omega$, after the quench of the electric field from 0 to 7.5 $\hbar \omega$, for the medium ($\gamma = 1\, \hbar \omega$) and large ($\gamma = 3\, \hbar \omega$) spin-rotation coupling strengths, respectively. The corresponding magnetization diagrams are presented in panels (a) and (b) of Fig.~\ref{fig:phase_diagrams}. The initial state of the system, $\ket{\Psi_0}$, is antisymmetric with $J=1$, $M_J=1$, and $M_S=-1$. It is composed predominantly of the $\ket{0}$ harmonic trap state. Firstly, we see that the amplitude of $\langle \hat{S}_z (t)\rangle$ variation increases with the spin-rotation strength. For a medium $\gamma$, the $\langle \hat{S}_z \rangle$ changes by around 20\% of its value, while for a large $\gamma$ -- by 100\%. As already stated, the $\langle \hat{S}_z \rangle$ value would be constant without the spin-rotation coupling.

Secondly, the DFT in insets of panels (a) and (b) of Fig.~\ref{fig:Mgquench} indicates that the dynamics is governed by a manifold of couplings. They result from the intermolecular interaction mixing the harmonic levels, the magnetic field lifting the degeneracy with respect to the projection of the total spin angular momentum, $M_S$, the electric field mixing the rotational states, and finally the spin-rotation coupling, which mixes states with different $J, S, M_J$, and $M_S$. The quench aims to assess the spin-rotation coupling strength, $\gamma$. Therefore we compare results from panels (a) and (b). While the number of present couplings is vast, most of them are negligible, especially in panel (a). Their impact on the dynamics grows with the spin-rotation coupling strength. Therefore, while just two couplings dominate the time evolution of $\langle \hat{S}_z \rangle$ for a medium $\gamma$, they are joined by many new ones for a large $\gamma$. In both cases, the dominant coupling is between the states taking part in the magnetization change, described in Sec.~\ref{sssec:Mg_spins12}. The strength of this coupling grows with the spin-rotation coupling strength~$\gamma$.

Summing up, in the case of the quench of the electric field, there are two signatures of the spin-rotation coupling strength in the time evolution of the magnetization, $\langle \hat{S}_z \rangle$. First is the size of $\langle \hat{S}_z (t)\rangle$ amplitude, which increases with $\gamma$. Second is the number of couplings present in the system and the amplitude of the dominant coupling, increasing with $ \gamma $.

Instead of quenching the electric field strength, the magnetic field can also be suddenly turned on. Panel~(c) of Fig.~\ref{fig:Mgquench} shows the time evolution of $\langle \hat{S}_z\rangle$ of the studied molecular system under the influence of the constant electric field of $5 \, \hbar \omega$, after the quench of the magnetic field from 0 to $4\, \hbar \omega$. The initial state, $\ket{\Psi_0}$, is already impacted by the constant electric field. It is predominantly $\ket{\Psi_0} = \ket{n=0, \, J=0, \, j_1=j_2=0, \, S = 0, \, M_S = 0}$, but mixed with the symmetric rotational state with $j_1$ and $j_2$ equal to 0 and 1. It also has a significant contribution from the higher harmonic states ($n=2,4$) due to the intermolecular interaction. The selected quench of the magnetic field does not modify the initial state significantly, so in the end we probe only couplings between $\ket{\tilde{\Psi}} \approx \ket{\Psi_0}$ and other eigenstates of the final Hamiltonian. This significantly limits the number of couplings influencing the dynamics, what is visible when comparing the corresponding DFTs in Fig.~\ref{fig:Mgquench}. Moreover, the initial state has $\langle \hat{S}_z \rangle = 0$, therefore the only significant couplings are between $\ket{\tilde{\Psi}}$ and the eigenstates with $\langle \hat{S}_z \rangle \neq 0$, which further limits the number of visible peaks in the DFT. However, the remaining peaks are related to the spin-rotation coupling as it is the only part of the Hamiltonian mixing states with different $M_S$.

In the magnetic field's quench, the initial state $\ket{\Psi_0}$ has the largest overlap with the fourth excited state of the final Hamiltonian, instead of the ground state, as it is in the electric field's quench. This means that the couplings governing the dynamics are not between the states taking part in the system's magnetization change. Moreover, the amplitude of $\langle \hat{S}_z (t)\rangle$ variation is not anymore linearly dependent on the spin-rotation coupling strength. It grows with the spin-rotation coupling strength till $\gamma$ reaches medium values, and then  remains almost constant. Different sets of couplings govern these two regimes. The first regime (small $\gamma$, $\langle \hat{S}_z (t)\rangle$ amplitude $\propto \gamma$) is dominated by a single coupling between $\ket{\tilde{\Psi}} \approx \ket{\Psi_0}$ and the nearest antisymmetric eigenstate with $J = M = 1$, $M_S = -1$, and $n=2$. This coupling strength grows with $\gamma$ till $\gamma$ reaches medium values. In the second regime (larger $\gamma$, constant $\langle \hat{S}_z (t)\rangle$ amplitude), the mentioned coupling strength decreases when $\gamma$ grows and a new coupling,  with the ground state of the quenched system, grows. Competition between two couplings results in the $\langle \hat{S}_z (t)\rangle$ amplitude being nearly independent of the spin-rotation strength. This quench scenario is thus less straightforward to determine the spin-rotation coupling strength value than the quench of the electric field. It still allows us to determine $\gamma$ by fitting the theoretical model to the experimental data.

The quench of the magnetic field strength proves useful also in determining whether the anisotropic part of the intermolecular interaction dominates the system's properties. As discussed in the Sec.~\ref{sssec:Mg_spins12}, in the system with medium interaction anisotropy, the main source of coupling between states with different $M_S$ values is the combination of the electric field and the spin-rotation coupling. When the electric field is missing, the main source of such a coupling is the large anisotropic part of the intermolecular interaction which brings states with higher rotational angular momenta to lower energies (see Fig.~\ref{fig:magnetisation_1/2}(c)).
It means that if the quench scenario from Fig.~\ref{fig:Mgquench}(c) is performed without the constant electric field, the nontrivial time evolution of $\langle \hat{S}_z \rangle$ indicates that the anisotropic part dominates the intermolecular interaction.

%------------------------------------------------------------------------------
\section{Conclusions}
\label{sec:summary}
%------------------------------------------------------------------------------
Within this work, we have studied the magnetic properties of two interacting ultracold polar and paramagnetic molecules in a one-dimensional harmonic trap. We have focused on the interplay of the molecular electronic spins, electric dipole moments, rotational structures, external electric and magnetic fields, and spin-rotation coupling. We have shown that control over the molecular system's magnetization could be achieved using an external electric field. This result is a complementary extension of the analogous studies focused on the free-space collisions. We have also presented the resulting magnetization diagrams depending strongly on two molecular properties of the system, namely the spin-rotation coupling and the anisotropic part of the intermolecular interaction. Motivated by the theoretical and experimental challenges in determining such molecular properties of few-body systems, we have employed the quench dynamics to find signatures of the anisotropic intermolecular interaction strength and the electronic spin-rotation coupling.

Our findings can be summarized as follows:

\begin{itemize}
\item The magnetization of the system can be controlled via external fields. The main underlying mechanism is the competition between the Zeeman and Stark effects. The spin-rotation coupling strength affects the smoothness of the transition between possible magnetization values.

\item The number of accessible magnetization values depends on selected $M_\text{tot}$ of the system, the electronic spins of the molecules, and the strength of the anisotropic part of the intermolecular interaction as it brings the states with higher total rotational momenta to lower energies.

\item The time evolution of the system's cloud size after the quench of the intermolecular interaction has clear signatures of the ratio between the anisotropic, $g_{\pm 1}$, and isotropic, $g_{0}$, part of the interaction. In the regime of large $g_{0}$ or $g_{0} = g_{\pm 1}$, the dynamics is governed by couplings between evenly separated harmonic states of the system. For $g_{\pm 1} > g_{0}$, the ladder of additional couplings becomes visible in the Fourier transform of the time evolution, coming from the antisymmetric ground state of the system. This ground state is highly sensitive to $g_{\pm 1}$ and may be used to determine its strength compared to $g_0$.

\item The time evolution of the magnetization after the electric field's quench depends strongly on the spin-rotation coupling strength. The larger spin-rotation coupling, the larger is the amplitude of the magnetization variation and the larger number of couplings governing the dynamics. It can thus be used to assess the strength of the spin-rotation coupling in the molecular system.

\item The time evolution of the magnetization after the magnetic field's quench is governed by a smaller number of couplings than after the electric field's quench. In the studied example, it is caused by a large similarity of the initial state to one of the eigenstates of the system after the quench. The dynamics probes then only couplings to this single eigenstate. While it may allow to assess the spin-rotation coupling strength, this scenario serves better for probing the anisotropic part of the intermolecular interaction.

\end{itemize}

The presented intrinsic coupling between the electric and magnetic properties of the studied model system paves the way towards studying the controlled magnetization of the ultracold many-body molecular systems trapped in optical tweezers or optical lattices.
The results provide also the first step in studying dynamical magnetic properties of a few-body molecular systems with varied geometries.
The potential applications range from quantum simulations of molecular multichannel many-body Hamiltonians to quantum information storing.

The studied model can be extended by including the fermionic or bosonic statistics of indistinguishable molecules or allowing dimers to be different.  Another direction is to incorporate the state dependence of molecular characteristics and trapping potential. The interaction potential with more realistic dependence on the relative distance between molecules may capture the physics of four-atom complexes that are now of central interest for ultracold molecular experiments. Another extension is the more realistic quench dynamics taking into account all correlations and dependencies between molecular characteristics. A natural extension to the many-body limit is the double molecular Mott insulator in an optical lattice with two molecules per site. The present system constitutes exotic monomers for such a system with large total rotational angular momenta in the ground state and magnetization controllable with the electric field.

%------------------------------------------------------------------------------
\begin{acknowledgments}
We would like to thank Dariusz Wiater for useful discussions.
We acknowledge the financial support from the Foundation for Polish Science within the Homing and First Team programmes co-financed by the EU Regional Development Fund and the PL-Grid Infrastructure.
A.D. acknowledges the financial support support from ERC AdG NOQIA, Spanish Ministry of Economy and Competitiveness (“Severo Ochoa” program for Centres of Excellence in R\&D (CEX2019-000910-S), Plan National FISICATEAMO and FIDEUA PID2019-106901GB-I00/10.13039 / 501100011033, FPI), Fundació Privada Cellex, Fundació Mir-Puig, and from Generalitat de Catalunya (AGAUR Grant No. 2017 SGR 1341, CERCA program, QuantumCAT U16-011424, co-funded by ERDF Operational Program of Catalonia 2014-2020), MINECO-EU QUANTERA MAQS (funded by State Research Agency (AEI) PCI2019-111828-2 / 10.13039/501100011033), EU Horizon 2020 FET-OPEN OPTOLogic (Grant No 899794), and the National Science Centre, Poland-Symfonia Grant No. 2016/20/W/ST4/00314.
\end{acknowledgments}
%------------------------------------------------------------------------------

\onecolumngrid

%------------------------------------------------------------------------------
\appendix
\section{Derivation of the size of the molecular cloud in the used basis}
\label{app:r2}
%------------------------------------------------------------------------------
The wave function of the studied molecular system is following:
\begin{equation}\label{eqA:psi}
|\Psi_k\rangle=\sum_{\substack{n,J,M,j_1,j_2,\\ S, M_S, s_1, s_2}}C^k_{\substack{n,J,M,j_1,j_2,S, M_S, s_1, s_2}} |n\rangle|J,M,j_1,j_2\rangle |S, M_S, s_1, s_2\rangle\,,
\end{equation}
where
\begin{equation}\label{eqA:basis}
\begin{split}
|n\rangle = \frac{1}{\sqrt{2^n n!}} \pi^{-1/4} \exp{\frac{-z^2}{2} H_n (z)} \,, \\
|J,M,j_1,j_2\rangle=\sum_{m_1,m_2}\langle j_1,m_1,j_2,m_2 |J,M\rangle |j_1,m_1\rangle|j_2,m_2\rangle\,, \\
|S, M_S, s_1, s_2\rangle = \sum_{m_{s_1},m_{s_2}}\langle s_1,m_{s_1},s_2,m_{s_2} |S,M_S\rangle |s_1,m_{s_1}\rangle|s_2,m_{s_2}\rangle\,.
\end{split}
\end{equation}
where $H_n$ are the Hermite polynomials, $\langle j_1,m_1,j_2,m_2 |J,M\rangle$ and $\langle s_1,m_{s_1},s_2,m_{s_2} |S,M_S\rangle$ are the Clebsch-Gordan coefficients, while $|j_i,m_i\rangle$ and $|s_i,m_{s_i}\rangle$ are the eigenfunctions of the rotational and spin angular momenta of the molecule $i$. The size of the molecular cloud is then (with $\beta \equiv J,M,j_1,j_2,S, M_S, s_1, s_2$):

\begin{equation}
\langle \hat{r}^2 \rangle = \bra{\Psi_k} \hat{r}^2 \ket{\Psi_{k}} =
\sum_{\substack{n,n',\beta}} C^k_{\substack{n,\beta}} C^k_{\substack{n', \beta}} \bra{n} \hat{r}^2 \ket{n'} = \sum_{\substack{n,n',\beta}} \frac{C^k_{\substack{n,\beta}} C^k_{\substack{n', \beta}}}{\sqrt{2^{n+n'} n! n'!}} \pi^{-\frac{1}{2}} \int_0^\infty dr \, r^2 e^{-r^2} H_n(r) H_{n'}(r)
\end{equation}

\begin{equation}
\begin{split}
\int_0^\infty dr \, r^2 e^{-r^2} H_n(r) H_{n'}(r) \stackrel{\text{\circled{1}}}{=} n! n'! \sum_{N=0}^{\min{n,n'}} \frac{2^N}{(n-N)! (n'-N)! N!} \int_0^\infty dr \, r^2 e^{-r^2} H_{n+n'-2N} (r) \\
\stackrel{\text{\circled{2}}}{=} n! n'! \sum_{N=0}^{\min{n,n'}} \frac{2^N (n+n'-2N)!}{(n-N)! (n'-N)! N!} \sum_{k=0}^{\text{floor}(\frac{n+n'}{2}-N)} \frac{(-1)^k 2^{n+n'-2N-2k-1} \Gamma (\frac{n+n'-2N+3}{2} - k)}{k! (n+n'-2N-2k)!}
\end{split}
\end{equation}

Hermite polynomials' properties used in calculations:
\begin{enumerate}[label=\protect\circled{\arabic*}]
\item $H_m(z) H_n(z) = m! n! \sum_{N=0}^{\min{m,n}} \frac{2^N H_{m+n-2N} (z)}{(m-N)! (n-N)! N!}$ \cite{Watson38},
\item $\int_0^\infty t^2 e^{-t^2} H_n(t) dt = n! \sum_{k=0}^{\text{floor}(\frac{n}{2})} \frac{(-1)^k 2^{n-2k-1} \Gamma (\frac{n+3}{2} - k)}{k! (n-2k)!}$ \cite{Mathematica}.
\end{enumerate}

\section{Spin-rotation coupling matrix elements in the used basis}
\label{app:spinrot}
%------------------------------------------------------------------------------
Here, we provide matrix elements of the spin-rotation component of the Hamiltonian given by Eq.~\eqref{eq:tot_ham} and \eqref{eq:partial_hams} in the computation basis of $|n,J,M_J,j_1,j_2,S,M_S,s_1,s_2\rangle\equiv|n\rangle|J,M_J,j_1,j_2\rangle |S, M_S, s_1,s_2\rangle$ as described in Sec.~\ref{sec:model}.

\begin{equation}
\begin{gathered}
\langle n,J,M_J,j_1,j_2,S,M_S,s_1,s_2 | \hat{H}_{\text{spin--rot}} | n',J',M_J',j_1',j_2',S',M_S',s_1',s_2' \rangle = \\
 \delta_{n n'} \delta_{M_{\text{tot}}, M_{\text{tot}}'} \delta_{j_1 j_1'} \delta_{j_2 j_2'} \delta_{s_1 s_1'} \delta_{s_2 s_2'} \\
\times\sum_{m_1 = -j_1}^{j_1} \sum_{m_2 = -j_2}^{j_2} \sum_{m_1' = -j_1}^{j_1} \sum_{m_2' = -j_2}^{j_2} \braket{j_1 m_1' j_2 m_2'}{J M_J} \braket{j_1 m_1' j_2 m_2'}{J' M_J'} \\
\times\sum_{m_{s_1} = -s_1}^{s_1} \sum_{m_{s_2} = -s_2}^{s_2} \sum_{m_{s_1}' = -s_1}^{s_1} \sum_{m_{s_2}' = -s_2}^{s_2} \braket{s_1 m_{s_1}' j_2 m_2'}{S M_S} \braket{s_1 m_{s_1}' s_2 m_{s_2}'}{S' M_S'} \\
\times\left(\delta_{m_1 m_1'} \delta_{m_2 m_2'} \delta_{m_{s_1} m_{s_1}'} \delta_{m_{s_2} m_{s_2}'} \gamma (m_1 m_{s_1} + m_2 m_{s_2}) \right. \\
+ \frac{\gamma}{2}\left(\delta_{m_{s_1}+1,m_{s_1}'} \delta_{m_{s_2},m_{s_2}'} \delta_{m_1-1,m_1'} \delta_{m_2 m_2'} + \delta_{m_{s_1},m_{s_1}'} \delta_{m_{s_2}+1,m_{s_2}'} \delta_{m_1 m_1'} \delta_{m_2 - 1,m_2'} \right.\\
\left.\left.+ \delta_{m_{s_1}-1,m_{s_1}'} \delta_{m_{s_2},m_{s_2}'} \delta_{m_1+1,m_1'} \delta_{m_2,m_2'} + \delta_{m_{s_1},m_{s_1}'} \delta_{m_{s_2}-1,m_{s_2}'} \delta_{m_1 m_1'} \delta_{m_2 + 1,m_2'} \right)\right)\,,
\end{gathered}
\end{equation}

\twocolumngrid
%\bibliography{quantum_magnetism_bib}

\end{document}